\documentclass[a4paper, 10pt]{article}

\usepackage{times,xcolor,amsmath,amssymb,epsfig,graphics,graphicx}
\usepackage{graphicx}
\usepackage{subfigure}

\newcommand{\arccot}{\textrm{ arccot }}

\newcommand{\e}{\textrm { e}}

\newcommand{\be}{\begin{eqnarray}}
\newcommand{\ee}{\end{eqnarray}}

\begin{document}

\begin{center}
{\Large \bf Heavy quarkonia properties  from a hard-wall confinement potential model with conformal symmetry perturbing effects}
\end{center}

\vspace{0.21cm}

\begin{center}
{\large Ahmed Al-Jamel}\\
Physics Department, Al Al-Bayt University, Mafraq 25113, Jordan \\
Email: aaljamel@gmail.com or aaljamel@aabu.edu.jo
\end{center}

\date{\today}  

\begin{flushleft}
{\bf Abstract:} Heavy $c\bar c$ and $b\bar b $ quarkonia are considered as systems confined within a hard-wall potential shaped after a linear combination of a cotangent-- with a square co-secant function. Wave functions and energy spectra
are then obtained in closed forms in solving by the Nikiforov-Uvarov method
the associated radial Schr\"{o}dinger equation  in the presence of a centrifugal term.  The interest in this potential is that 
in one parametrization it can account for a conformal symmetry of the strong interaction, and in another for  its perturbation, a reason for which we here employ it to study status of conformal symmetry in the heavy flavor sector. The resulting predictions on  heavy quarkonia mass spectra and root mean square radii  are compared with the available experimental data, as well as with predictions by  other theoretical approaches. We observe that a relatively small conformal symmetry perturbing term in the potential suffices to achieve good agreement with data. 

\end{flushleft}

\section{Introduction} 
Following the remarkable recent discovery of the Higgs boson at the Large Hadron Collider (LHC),  further open questions rapidly moved to the focus of the experimental projects of this leading experimental facility, among them the study of the production mechanisms  of heavy quarkonia (bound states of heavy quark and heavy anti-quark),
 a process whose exploration  is expected to provide significant new insights into our understanding of the strong interaction dynamics
(for details one can refer to \cite{Li01},\cite{CERN2005},\cite{Brambilla2011},\cite{Andronic16} and references therein). Because of their heavy masses ($m_q \gg\Lambda_{QCD}$),  
 the dynamical description of these systems grows  more complex. The field theoretical perturbative methods  may not always be efficient in that regard, 
 a circumstance that can be remedied to some extent by the employment of effective degrees of freedom such as constituent quarks, interacting via 
 non-relativistic  potentials. 
 In the last two decades, several  quark potential models have been worked out for the sake of 
 data evaluation on heavy quarkonia, which are based on interactions such as the Cornell and  Martin potentials, the  Cornell plus quadratic potential, 
 among others. 
 The potentials used in the aforementioned models reflect to some extent the two basic features of strong interaction, asymptotic freedom, and confinement, predominantly soft walled. 
However, none of them relates to  the conformal $SO(2,4)$ symmetry of strong interaction.  The latter is best established at the  kinematic level in the regime of the asymptotic freedom where the running coupling becomes practically vanishing, causing that the constituent quarks start gradually loosing  their gluon  ``dressings'', evolving  to partons, the free matter fields in the QCD Lagrangians.  The parton masses in the unflavored sector are sufficiently small and allow conformal symmetry to acquire importance there. Moreover, recently experimental hints have been obtained also on possible viability of a dynamical  conformal symmetry in the infrared regime through the observed  walking of the strong coupling towards a fixed value at origin \cite{Andre}. For the sake of description of this phenomenon a quark potential is required that captures the features of the conformal symmetry, in addition to confinement and asymptotic freedom. A properly parameterized linear combination of a cotangent- and a squared co-secant functions, known under the name of the ``trigonometric Rosen-Morse potential'' \cite{Rosen32}, also managed by the super-symmetric quantum mechanics, has been reported in \cite{Kirchbach2016} to serve this purpose. In the latter work, the unflavored mesons have been studied and a fairly well agreements between predictions and data  on the spectra could be found. For the heavy flavors, one expects the conformal symmetry to be violated by the large masses of the charmed and bottom quarks, and it is an intriguing task to identify  a potential parameter as a signature for such a violation.  The interaction under discussion is in addition a hard walled potential in so far as it is of an infinite depth, and the associated wave functions are vanishing in the outer region. 
The goal of the present work is twofold. On the one side, we wish to explore the effect of the hard walls on the properties of the heavy $(c\bar c)$ and $(b\bar b)$ quarkonia,  and on the other, to  shad some light on the status of conformal symmetry in this sector. 
In order to account for the second effect, 
we here extend the trigonometric Rosen-Morse potential applied in \cite{Kirchbach2016} to the description of a good dynamical conformal symmetry by a properly designed term that triggers  the perturbation of the latter.

 Before proceeding further, we like to notice that in being hard walled,
 the  potential under discussion, has frequently found applications to the  studies of  systems featuring confinement and  ranging from quantum dots \cite{Kurochkin}, over Coulomb fluids \cite{Gaillol},   
to hadron structure \cite{Kirchbach2016},\cite{Kirchbach2018}.  Furthermore, in \cite{Ciftci13}, the same potential has been considered as belonging to  
a  broader class  of trigonometric interactions,
 to which  exact and approximate solutions  have been found by means of the asymptotic iteration method. In addition,  in \cite{Ahmed10},  the trigonometric 
 Rosen-Morse potential and its exact solutions have served as a point  of departure towards the constructions of  new  potentials by subjecting it to transformations for which bound states spectra and the corresponding wave functions have been calculated. Finally, in \cite{Jia13} the $q$ deformed Rosen-Morse potential (in its hyperbolic version) 
has been defined  and used as an improved model to the Tietz potential in  investigations of diatomic molecules.

We here employ a version of the trigonometric Rosen-Morse potential 
 within the framework  of non-relativistic quantum mechanics to generate
 heavy quarkonia spectra and wave functions in three dimensional space.

The article is structured as follows. In the next section we  
solve the potential problem of interest by the aid of the Nikiforov-Uvarov method. 
Comparison with available experimental data and related theoretical calculation  are  presented in section 3. The text closes with a brief summary.

\section{Heavy quarkonia potential problem  and solutions}
One of the most popular techniques for solving elliptic differential equations, as are the stationary  Schr\"odinger equations, has been developed by Nikiforov and Uvarov \cite{Nikiforov1988}, and will  be abbreviated in what follows by NU. 
Its most essential technical aspects are briefly highlighted in the subsequent section, for the sake of self sufficiency of the presentation.

\subsection{The Nikiforov-Uvarov method} 

The method is based on the assumption that the stationary one-dimensional Schr\"{o}dinger equation can be cast in the form of the generalized hyper-geometric equation as,
\begin{equation}
\label{eq1} 
\psi ^{\prime \prime }(s)+\frac{\tilde{\tau}(s)}{\sigma (s)}\psi ^{\prime }(s)+\frac{\tilde{\sigma}(s)}{\sigma ^{2}(s)}\psi(s)=0,
\end{equation}
by the aid of some suitable point-canonical transformation. Here,  $\tilde{\tau}(s)$ is a polynomial of at most first degree, while $\sigma(s)$ and $\tilde{\sigma}(s)$ are polynomials of at most second degree \cite{Nikiforov1988}. 
Utilizing factorization, 
\begin{equation}
\label{eq2}
\psi(s)=\phi(s)y(s),
\end{equation}
Eq.(\ref{eq1}) allows for a transformation to an equation of the standard hyper-geometric type:
\begin{equation}
\label{eq3} 
\sigma(s)y^{\prime \prime }+\tau(s)y^{\prime }+\lambda y=0,
\end{equation}
where  $\tau (s)$ has been defined as
\begin{equation}
\label{eq10}
\tau (s)=\tilde{\tau }(s)+2\pi (s),
\end{equation}
while the $\pi (s)$ function is given by,
\begin{equation}
\label{eq7}
\pi(s)  =\frac{\sigma ^{\prime}(s)-\tilde{\tau}(s)}{2}\pm
\sqrt{\left(\frac{\sigma^{\prime}(s)-\tilde{\tau}(s)}{2}\right)^{2}-
\tilde{\sigma}(s)+k{\sigma}(s)}.
\end{equation}
The latter is supposed to be a polynomial of first degree at most, 
so that  the expression under the square root in Eq.(\ref{eq7}) could take the shape of  a square of a polynomial of first degree. The parameters $\lambda$ and  $k$ then satisfy the condition,
\begin{equation}
\label{eq8}
\lambda =k+\pi ^{\prime}(s).
\end{equation}

Provided the sign of the derivative of the $\tau(s)$ function is negative, the equation (\ref{eq3}) can be shown to have polynomial solutions $y_{n}(s)$, generated by the Rodrigues formula according to,
\begin{equation}
\label{eq5}
y_{n}(s)=\frac{B_{n}}{\rho (s)}\frac{d^{n}}{ds^{n}}\left[ \sigma ^{n}(s)\rho (s)\right].
\end{equation}
Here, $B_{n}$ is a normalization constant, and $\rho(s)$, termed to as density-, or weight function,  must satisfy the condition
\begin{equation}
\label{eq6}
 (\sigma (s) \rho (s) )^{\prime }=\tau (s)\rho (s).
\end{equation}
Moreover, for this case one finds an $n$ dependent $\lambda$ according to,
\begin{equation}
\label{eq9}
\lambda _{n}=-n\tau ^{\prime 
}(s)-\frac{n(n-1)}{2}\sigma ^{\prime \prime }(s),~~~n=0,1,2,...,
\end{equation}
With that, the function $\phi(s)$ is found to satisfy
\begin{equation}
\label{eq4}
\frac{\phi (s)^{\prime }}{\phi (s)}=\frac{\pi(s)}{\sigma (s)}.
\end{equation}
Equating Eq.(\ref{eq8}) with Eq.(\ref{eq9}), it can be shown to allow to obtain  the energy eigenvalues of the Schr\"odinger equation.
 
\subsection{Solving the Schr\"odinger equation with the trigonometric Rosen-Morse potential by the aid of the NU method}
We are interested in investigating quantum systems confined in the  trigonometric Rosen-Morse potential, to be abbreviated by ``tRM'',  of the form given by  \cite{Kirchbach2016},\cite{Kirchbach2018} (here in dimensionless units)
\begin{equation}
\label{cot}
V\left(\frac{r}{a}\right)=-V_0 \cot \left(\frac{r}{a}\right)+d (d+1) \csc ^2\left(\frac{r}{a}\right) +a^2\frac{\ell (\ell +1)}{r^2} , ~~~~0<r<a\pi, 
\end{equation}
with $\ell$ standing for the angular momentum value.  In the approximation,
\begin{equation}
 V\left(\frac{r}{a} \right)\approx -
V_0\frac{a }{r} +V_0\frac{r}{3a} + \left[ \ell (\ell +1) +d(d+1)\right]\frac{a^2}{r^2},
\label{Petra}
\end{equation}
 it reproduces the functional form of a potential of a frequent use \cite{Rani18} in the spectroscopy of quarkonia. 
This approximation emerges from keeping the lowest terms in the series expansions of, 
 \begin{equation}
\label{cot-approx}
- \cot\left(\frac{r}{a}\right)\approx -\frac{a}{ r}+\frac{1}{3} \frac{r}{a}
, 
\end{equation}
and
\begin{equation}
\label{csc-approx}
\csc^2\left(\frac{r}{a}\right)\approx \frac{1}{\frac{r^2}{a^2}}-\frac{1}{15}
\frac{r^2}{a^2},
\end{equation}
respectively.
Within this context, the potential in (\ref{cot}) can be viewed as an upgrade of the one in (\ref{Petra}).
The stationary  Schr\"{o}dinger equation we are interested in here reads
\begin{eqnarray}
{\mathcal H}_{tRM}\left(\frac{r}{a} \right)u_{n \ell}\left(\frac{r}{a}\right )&=&E
u_{n \ell}\left(\frac{r}{a}\right ), \nonumber\\
{\mathcal H}_{tRM}\left(\frac{r}{a} \right)&=&-\left[\frac{\hbar^2}{2\mu}\frac{d^2}{dr^2}+\frac{\hbar^2}{2\mu a^2}\left[ V_0 \cot \left(\frac{r}{a}\right)-
d (d+1) \csc ^2\left(\frac{r}{a}\right)
-\frac{\ell(\ell+1)}{2\mu r^2}\right]\right]. 
\label{Rania}
\end{eqnarray}
 With the aim to arrive at exact solutions, we shall approximate the centrifugal term in the latter equation as,
\begin{equation}
\label{PetraJ}
\frac{1}{r^2}\approx  \frac{1}{a^2\sin^2(\frac{r}{a})}\equiv\frac{1}{a^2}\csc^2\left(\frac{r}{a}\right), ~~~~\frac{r}{a}<<1.
\end{equation}
In effect, the wave equation to be solved takes in dimensionless units the shape,
\begin{equation}
\left[
a^2\frac{d^2}{dr^2}+\frac{2\mu c^2 a^2}{\hbar^2 c^2} E+V_0 \cot 
\left(\frac{r}{a}\right)-d (d+1) \csc ^2\left(\frac{r}{a}\right)-
\frac{\ell(\ell+1)}{ \sin^2 \left(\frac{r}{a} \right)}
\right]u_{n \ell}\left(\frac{r}{a}\right )=0.
\label{RM1}
\end{equation}

Within the above context, the equation (\ref{RM1})   
can be interpreted in a twofold way.

\begin{itemize}
\item Without  the approximation in (\ref{PetraJ}) it represents 
the radial part  of the flat-space Such\"odinger equation with the trigonometric Rosen-Morse potential in the presence of a centrifugal term.
\item With this approximation, same equation can be read as equation in the angular variable, $\chi=r/a$, where $a$ plays the role of a spherical radius, while $r$ acquires meaning of the arc on a great circle, read off from the ``North pole''. The fact is that  the term $\ell (\ell +1)\csc^2\chi$ represents the centrifugal barrier on the three dimensional sphere, $S^3$, whose isometry $SO(4)$, is the maximal compact subgroup of the conformal group, $SO(2,4)$, a reason for which the equation for $d=0$ describes dynamical conformal symmetry. The additional  $d(d+1)\csc^2(r/a)$ term could be interpreted within this context  as a conformal symmetry perturbation that alters the shape of the wave functions (see discussion to the end of section 3 below).  
\end{itemize}

Now, with the help of $\csc^2(\frac{r}{a})=1+\cot^2(\frac{r}{a})$ and the substitutions
\begin{equation}
\label{34}
\epsilon=\frac{2\mu}{\hbar^2}E,~~~V_c=\frac{\hbar^2 V_0}{2\mu a^2},~~~U_0=\frac{V_0}{a^2},~~~\gamma=\frac{1}{a^2}\left(d (d+1)+\ell(\ell+1)\right)
\end{equation}
where $V_c$ is the amplitude of the cotangent part of the potential in units of MeV, while the rest of variables are in units of fm$^{-2}$, Eq.~(\ref{RM1}) becomes
\begin{equation}
\label{33}
\left[\frac{d^2}{dr^2}+\left(\epsilon+U_0 \cot \left(\frac{r}{a}\right)-\gamma-\gamma\cot^2\left(\frac{r}{a}\right)\right)\right]u_{n \ell}\left(\frac{r}{a}\right)=0,
\end{equation}
Introducing a suitable transformation as $u_{n \ell}(r/a)=\sin(r/a)f(r/a)$, which ensures that $u_{n \ell}(0)=u_{n \ell}(\pi)=0$, followed by a change of variable, $y=\cot\left(\frac{r}{a}\right)$, and upon making use of the Ansatz \cite{Ciftci13}
\begin{equation}
f(y)=(1+y^2)^{\frac{\alpha}{2}} e^{-\beta\arccot(y)} g(y), ~~~-\infty <y< \infty,
\end{equation}
the equation ~(\ref{33}) is found to take the following shape,
\begin{equation}
\label{36a}
\frac{d^2g(y)}{dy^2}+\frac{2(\alpha y+\beta)}{1+y^2}\frac{dg(y)}{dy}+\frac{\left(\alpha(\alpha-1)-a^2\gamma\right)y^2+\left(2\beta(\alpha-1)+U_0 a^2\right)y+\alpha+a^2\epsilon+\beta^2-a^2\gamma -1}{(1+y^2)^2}g(y)=0.
\end{equation}
The freedom in the selection of $\alpha$ and $\beta$ in Eq.~(\ref{36a}) allows us to choose:
\begin{eqnarray}
\label{36b}
2\beta(\alpha-1)+U_0 a^2 &=&0. \nonumber\\
a^2 \epsilon-(\alpha-1)^2+\beta^2&=&0,
\end{eqnarray} 
then Eq.~(\ref{36a}) becomes
\begin{equation}
\label{36}
\frac{d^2g(y)}{dy^2}+\frac{2(\alpha y+\beta)}{1+y^2}\frac{dg(y)}{dy}+\frac{\left(\alpha(\alpha-1)-a^2\gamma\right)(1+y^2)}{(1+y^2)^2}g(y)=0.
\end{equation}
In making now use  of the equations (\ref{eq1})--(\ref{eq8}) from the  NU method, allows for the following identifications:  $\tilde{\tau}(y)=2(\alpha y+\beta)$, $\sigma(y)=1+y^2$, $\tilde{\sigma}(y)=\left(\alpha(\alpha-1)-a^2\gamma\right)(1+y^2)$. Therefore, we can proceed executing the further prescriptions of  the NU method. In so doing, we first find the function $\pi(y)$ in (\ref{eq7}) expressed as,
\begin{equation}
\label{eq37}
\pi (y)=(1-\alpha)y-\beta\pm\sqrt{k\left(y^2+1\right)-\left(\alpha(\alpha-1)-a^2\gamma\right) \left(y^2+1\right)+\left((1-\alpha ) y-\beta \right)^2}.
\end{equation}
The k-roots are $k_1=-\alpha +\alpha ^2-a^2\gamma$ and , $k_2=\alpha-\beta ^2-\gamma-1$. Therefore we have four possible choices for $\pi(y)$. The only choice that gives physically acceptable solutions is the one with $k_1$ corresponding to  negative sign  in Eq.~(\ref{eq10}). The results is $\pi(y)=0$. The associated function $\tau$ is then $\tau(y)=2\alpha y +2\beta$. This function has a
negative sign derivative only if $\alpha<0$. In this case we obtain:
 \begin{equation}
\label{eq38}
\alpha^2+(2n-1)\alpha+n(n-1)-a^2\gamma=0.
\end{equation}
Solving the latter equation  we obtain 
\begin{equation}
\alpha=-n+\frac{1}{2}\pm\frac{1}{2}\sqrt{4a^2\gamma+1}. 
\end{equation}
The constraint $\alpha < 0$ amounts to the following choice for $\alpha$:
\begin{equation}
\label{eq39}
\alpha=-n+\frac{1}{2}-\frac{1}{2}\sqrt{4a^2\gamma+1}, 
\end{equation}
which is always negative for any $n\geq 0$ and $\gamma \geq 0$. Substituting for $\gamma$ from Eq.~(\ref{34}), we obtain
\begin{equation}
\label{eq40}
\alpha=-n+\frac{1}{2}-\frac{1}{2}\sqrt{4d(d+1)+4\ell(\ell+1) +1}.
\end{equation}
Plugging this in Eq.~(\ref{36b}), we obtain the energy eigenvalues as
\begin{equation}
\label{eq41}
E_{n\ell}=\frac{\hbar^2}{2\mu a^2}
\left(-\alpha -1 \right)^2-\frac{\hbar^2 V^2_0}{8\mu a^2\left(-\alpha -1\right)^2}.
\end{equation}
Here $n,\ell=0,1,2,...$. Note that this energy formula admits as well positive as negative value of $V_0$ since this parameter enters the energy formula through its square, $V_0^2$. {}The special case of  $d=0$ amounts to,
\begin{equation}
\label{eq42}
E_{n\ell}=\frac{\hbar^2}{2\mu a^2}\left(n+\ell +1\right)^2-\frac{\hbar^2 V_0^2}{8\mu a^2\left(n+\ell+1\right)^2},
\end{equation}
and reproduces the related result in  \cite{Ciftci13,Kirchbach2018}. 
For small $d$, the expansion of Eq.~(\ref{eq41}) around $d=0$ yields
\begin{eqnarray}
\label{eq422}
E_{n\ell}&\approx& \frac{\hbar^2}{2\mu a^2}\left(n+\ell +1\right)^2-\frac{\hbar^2 V_0^2}{8\mu a^2\left(n+\ell+1\right)^2}\nonumber\\
&+& d \left(\frac{2\hbar^2 \left(\sqrt{\ell^2+\ell+1}+n+\frac{1}{2}\right)}{a^2 \mu  \sqrt{\ell^2+\ell+1}}+\frac{2\hbar^2 V_0^2}{a^2 \mu  \sqrt{\ell^2+\ell+1} \left(\sqrt{\ell^2+\ell+1}+2 n\right)^3}\right),
\end{eqnarray}
which will be used later.

For the corresponding eigenfunctions, one can easily verify that, using Eq.~(\ref{33}),  the following equations hold valid,
\begin{align*}
    &\phi(y)=constant, & \\
    &\rho(y)=(1+y^2)^{\alpha-1}e^{2\beta \arctan(y)} &\\ 
    &Y_n(y)=B_n e^{-2\beta \arctan(y)} (1+y^2)^{1-\alpha}\frac{d^n}{dy^n}\left((1+y^2)^{n+\alpha-1}e^{2\beta \arctan(y)}\right),&\\
\end{align*}
hold valid.
{}From that,  the Rodrigues formula that generates the polynomials $g_n(y)$ is concluded as
\begin{equation}
\label{43}
g_n(y)=B_n (1+y^2)^{1-\alpha} e^{-2\beta\arctan(y)}\frac{d^n}{dy^n}\left((1+y^2)^{n+\alpha-1}e^{2\beta\arctan(y)}\right)
\end{equation}
where $B_n$ are the normalization constants. Note that it depends on the $\arctan(y)$. According to this formula, and ignoring the normalization constants, we have
\begin{eqnarray}
g_0(y)&=&1,\\
g_1(y)&=&2( \alpha y+\beta),\\
g_2(y)&=&2 (\alpha +1) (2 \alpha +1) y^2+4 (2 \alpha  \beta +\beta )y+2 
\left(\alpha +2 \beta ^2+1\right),\\
g_3(y)&=&4 (\alpha +1) (\alpha +2) (2 \alpha +3) y^3+12 (\alpha +1) 
(2 \alpha +3) \beta  y^2+4  {\Big(}6 (\alpha +1) \beta ^2 \nonumber\\
&+&3 (\alpha +1)(\alpha +2){\Big)}y+4 {\Big(}3 \alpha  
\beta +2 \beta ^3+5 \beta {\Big)}\\
...&=&...
\end{eqnarray}
which coincide with the related expressions reported in \cite{Ciftci13}. These polynomials $g_n(y)$ are related to the known as Romanovski polynomials $R_n^{\beta,\alpha}(y)$, which are defined by the Rodrigues formula \cite{Raposo07}
\begin{equation}
\label{44}
R_n^{\beta,\alpha}(y)=\frac{1}{\omega^{\beta,\alpha}(y)}\frac{d^n}{dy^n}\left(\omega^{\beta,\alpha}(y) S(y)^n\right)
\end{equation}
where
\begin{equation}
\label{45}
\omega^{\beta,\alpha}(y)=(1+y^2)^{\alpha-1} \e^{-\beta\arccot(y)}
\end{equation}
is the weight function, equivalent to $\rho(y)$ in the NU method, and $S(y)=1+y^2$.
This can be seen if we use
\begin{equation}
\label{cot-tan}
\arctan(y)+\arccot(y)=\frac{\pi}{2},~~~Re[y]\geq 0
\end{equation}
and noting that 
\begin{eqnarray}
\label{45b}
\omega^{2\beta,\alpha}(y)&=&(1+y^2)^{\alpha-1} \e^{-2\beta\arccot(y)} \nonumber\\
&=& (1+y^2)^{\alpha-1} \e^{2\beta\arctan(y)-\beta\pi}
\end{eqnarray}
and thus
\begin{eqnarray}
\label{44b}
R_n^{2\beta,\alpha}(y)&=&(1+y^2)^{1-\alpha} \e^{-2\beta\arctan(y)-\beta\pi}\frac{d^n}{dy^n}\left((1+y^2)^{n+\alpha-1} \e^{2\beta\arctan(y)+\beta\pi}\right) \nonumber\\
&=& (1+y^2)^{1-\alpha} \e^{-2\beta\arctan(y)} \frac{d^n}{dy^n}\left( (1+y^2)^{n+\alpha-1} \e^{2\beta\arctan(y)}\right) \nonumber\\
&=&g_n(y) ~~\text{(up to a multiplicative constant)}
\end{eqnarray}
The orthonormalization of Romanovski polynomials is:
\begin{equation}
\label{46}
\int_{-\infty}^{\infty}R_n^{\beta,\alpha}(y)R_m^{\beta,\alpha}(y)\omega^{\beta,\alpha}(y)dy=B_n^2 \delta_{nm}, ~~~\text{iff}~~~m+n< 1-2\alpha
\end{equation}
which leads to the orthonormalization condition   
\begin{equation}
\label{47}
\int_{-\infty}^{\infty}g_n(y)g_m(y)(1+y^2)^{\alpha-1}e^{2\beta\arctan(y)}dy=B_n^2 \delta_{nm},~~~\text{iff}~~~m+n< 1-2\alpha.
\end{equation}
Thus, for the case when the polynomial's parameters do not depend on its degree, only a  finite number of  polynomials, namely those 
 satisfying the constraint $m+n< 1-2\alpha$, are orthogonal, a circumstance 
 known in literature under the name of `` finite orthogonality''.
However, for the case considered here, in which the polynomial parameters depend on the polynomial degrees in accord with the equations (\ref{eq40}) and 
$\beta$ from  (\ref{36b}), an infinite orthogonality is encountered \cite{Raposo07}. The reduced wave function can now be written in a closed form as a function of $r/a$ as
\begin{eqnarray}
\label{wf}
u_{n \ell}\left(\frac{r}{a}\right)&=&B_n\sin\left(\frac{r}{a}\right) f\left(\frac{r}{a}\right) \nonumber\\
&=&B_n\sin\left(\frac{r}{a}\right) \left(1+\cot^2\left(\frac{r}{a}\right)\right)^{\frac{\alpha}{2}} \e^{-\beta\arccot\left(\cot\left(\frac{r}{a}\right)\right)} g_n
\left(\cot \left(\frac{r}{a}\right) \right) \nonumber\\
&=&B_n\sin\left(\frac{r}{a}\right)\left(1+\cot^2\left(\frac{r}{a}\right)\right)^{1-\frac{\alpha}{2}} \e^{-2\beta\arctan\left(\cot\left(\frac{r}{a}\right)\right)-\beta\frac{r}{a}}\nonumber\\
&\times&\frac{d^n}{d\left(\cot\left(\frac{r}{a}\right)\right)^n}\left(\left(1+\cot\left(\frac{r}{a}\right)^2\right)^{n+\alpha-1}\e^{2\beta\arctan\left(\cot\left(\frac{r}{a}\right)\right)}\right).
\end{eqnarray}
where we have used Eq.~({\ref{44b}) and the fact $\cot^{-1}\left(\cot\left(\frac{r}{a}\right)\right)=\frac{r}{a}$. Here the value of $\beta$ is
\begin{equation}
\label{beta}
\beta=\frac{-V_0}{2(\alpha-1)}
\end{equation}
which must be positive to ensure the exponential fall-off of the wave functions needed for convergence of various integrals related to physical observable. This condition is already satisfied as $\alpha$ is found to be negative for the applicability of the method.
\section{Analyzing data on heavy quarkonia}

\subsection{Mass spectra}
The mass spectra of heavy quarkonia can be produced using the formula:
\begin{equation}
\label{eq48}
 M_{q\bar{q}}=m_q+m_{\bar{q}}+E_{n\ell},
\end{equation}
where $E_{n\ell}$ is given by  Eq.~(\ref{eq41}), $m_q$ and $m_{\bar{q}}$ are the masses of the constituent quarks, and anti-quarks, respectively. The reduced mass $\mu$ that appears in Eq.(\ref{eq41}) is defined in the standard way as $\mu=\frac{m_q m_{\bar{q}}}{m_q + m_{\bar{q}}}$. For $b\bar{b}$ and $c\bar{c}$ systems, we adopt the numerical values of these masses as $m_b =4.67$ GeV for bottomonium, and  $m_c =1.50$ GeV for charmonium. Then, the corresponding reduced masses are  $\mu_b =2.335$ GeV and $\mu_c =0.75$ GeV, respectively. The magnitude $V_0$ of the cotangent term in (\ref{cot}) has been parameterized as $V_0=\alpha_sN_c$ with $\alpha_s$ being the value of the strong coupling for $m_c$, and $N_c=3$ the number of colors. The values of $\alpha_s$ at the heavy quark masses of interest have been measured with varying precision and reported for example in \cite{Rani18}, \cite{Das16},\cite{Godfrey16},\cite{Eichten94},\cite{Maezawa16}. In the present study we opted in favor of  using in the numerical calculations a common reasonable  averaged value of $\alpha_s=0.2$. The potential parameters $a$ and $d$ can then be obtained by simultaneous fitting available experimental data on excited heavy flavor mesons. By adjusting  masses of $1S$ and  $2P$ states of  $b\bar{b}$ and  $c\bar{c}$ mesons, we extracted the values of these parameters for each one of the two sectors. Then, the remaining states in each one of the two  spectra are predicted using the obtained parameters according to  Eqs.~(\ref{eq48}) and (\ref{eq41}). The results are presented  in Tables~(\ref{Table1}-\ref{Table2}). These show that our energy formula can reproduce the masses in pretty good agreement with data.
In Figures~\ref{fig2}, we plotted the trigonometric Rosen-Morse  potential
with  the adjusted  parameters for the cases of an  $S$-wave--,  and a $P$ wave state for charmonium. It is visible how  the corresponding potentials bend strongly upwards from both sides of the interval, a behavior that reflects the strong confinement of the systems under study. The bending of the trigonometric Rosen-Morse potential for zero angular momenta is due to the $d(d+1)\csc^2(r/a)$ term and therefore to the conformal symmetry perturbation.  

\begin{table}[!htbp] \caption{Mass spectra for $c\bar{c}$ in GeV.According to \cite{Kirchbach2016} the magnitude $V_0$ of the cotangent term in (\ref{cot}) has been parameterized as $V_0=\alpha_s N_c$ with $\alpha_s$ being the value of the strong coupling for $m_c$, and $N_c=3$ the number of colors. Using $\alpha_s=0.2$, the dimensional magnitude emerges as $V_c=0.050$ GeV for the reduced mass of $\mu_c=0.75$ GeV and $m_c =1.5$ GeV.   
Furthermore, $d=0.109, a =2.822$ GeV$^{-1}=0.56$ fm. Experimental data are taken from \cite{PDG18,Olive14}.} 
\centering 
\begin{tabular}{ccccccc}
\hline
$nL$ & Present work  & Present work $d=0$ & Experimental & Ref.~\cite{Rani18}\\ \hline
1S & 3.097 & 3.076 & 3.097 & 3.096 \\ 
 2S & 3.371 & 3.333 & 3.686  & 3.686\\ 
1P & 3.347 & 3.333 & 3.525  & 3.214\\
3S & 3.809 & 3.752 & 4.040  & 4.275\\ 
2P & 3.773 & 3.752 & 3.773 & 3.773\\ 
1D & 3.765 & 3.752 & 3.770 & 3.412\\ 
4S & 4.414 & 4.339 & 4.263 & 4.865\\ 
3P & 4.366 & 4.339 & ---   & ---\\ 
2D & 4.355 & 4.339 & --- & ---\\ 
1F & 4.350 & 4.339 & --- & ---\\
\hline
\end{tabular}
\label{Table1}
\end{table}
\begin{table}[!htbp] \caption{Mass spectra for $b\bar{b}$ (in GeV). According to \cite{Kirchbach2016} the magnitude $V_0$ of the cotangent term in (\ref{cot}) has been parameterized as $V_0=\alpha_sN_c$ with $\alpha_s$ being the value of the strong coupling for $m_b$, and $N_c=3$ the number of colors. Using $\alpha_s=0.2$, the dimensional magnitude emerges as $V_c=0.059$ GeV for the reduced mass of $\mu_b=2.335$ GeV, ($m_b=4.67$ GeV).   
Furthermore, $d= 0.131, a =1.468$ GeV$^{-1}=0.290$ fm. Experimental data are taken from \cite{PDG18,Olive14}.}
\centering 
\begin{tabular}{cccccc}
\hline
$nL$ & Present work & Present work $d=0$   & Experimental & Ref.~\cite{Rani18}\\
\hline
1S & 9.460  & 9.430 & 9.460  & 9.460 \\ 
2S & 9.789  & 9.735 & 10.023  & 10.023\\ 
1P & 9.755  & 9.735 & 9.899   & 9.492\\ 
3S & 10.312 & 10.233 & 10.355 & 10.585\\ 
2P & 10.262 & 10.233 & 10.260   & 10.038\\ 
1D & 10.250 & 10.233 & 10.164  & 9.551\\ 
4S & 11.034 & 10.929 & 10.580 & 11.148\\ 
3P & 10.967 & 10.929 & ---   & ---\\ 
2D & 10.952 & 10.929 & --- & ---\\ 
1F & 10.945 & 10.929 & --- & ---\\
 \hline
\end{tabular}
\label{Table2}
\end{table}

\begin{figure}
\centering
\subfigure{\includegraphics[scale=0.80]{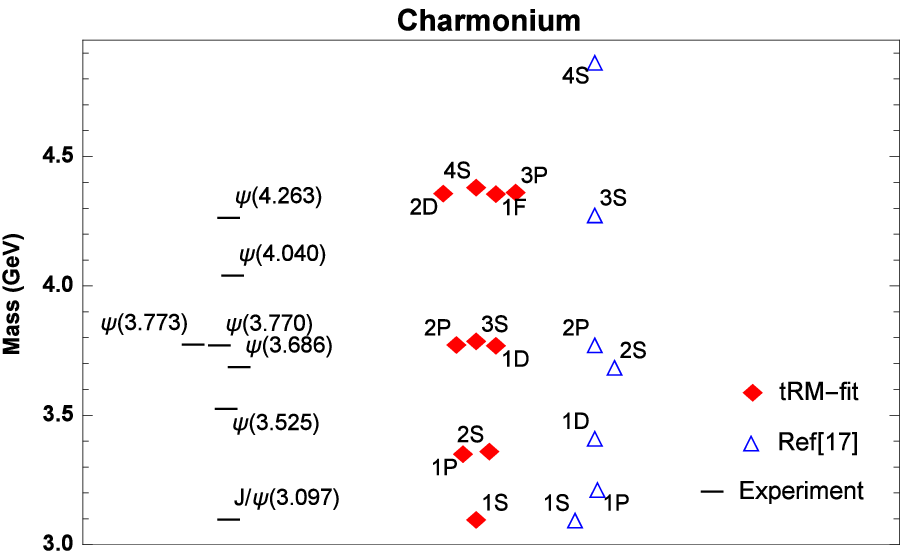}}
\subfigure{\includegraphics[scale=0.80]{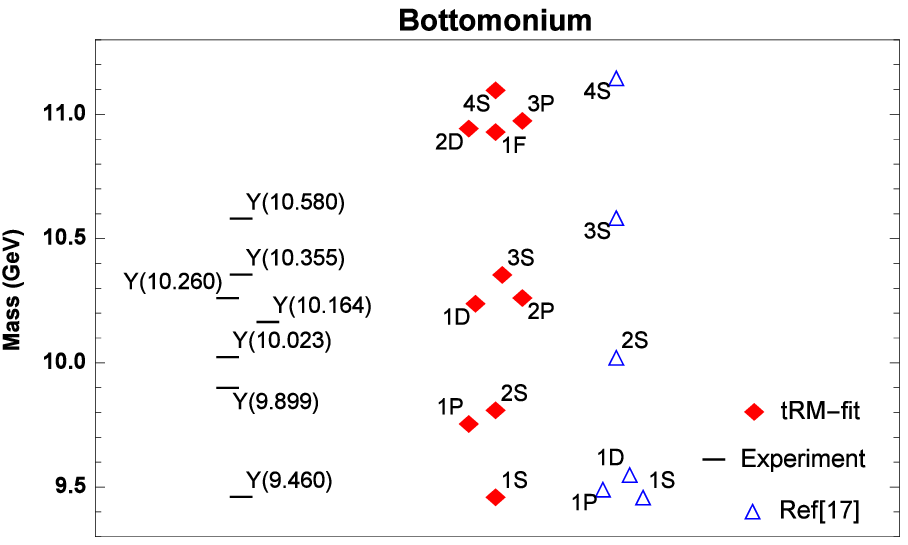}}
\caption{The predictions (red diamonds) of the charmonium (left panel) and bottomonium (right panel) spectra by the tRM potential. The mass in GeV is read off from the vertical axis, while the horizontal line is marginal. The experimental data are taken from \cite{PDG18,Olive14}. As a comparison we bring the theoretical predictions by \cite{Rani18} (blue triangles).
  }
\label{figc}
\end{figure}
The fits presented in the tables show a satisfactory  overall agreement between predictions and data. They also show that in each one of the cases the $d$ value is maintained surprisingly  small indeed, a reason for which  
the predicted level splittings strongly underestimate the experimentally observed ones, although the level orderings (the signs of the splittings) come out correct. In contrast, some of the level splitting predicted by \cite{Rani18}, like the $3S$--$1D$ splitting in the $b\bar b$ sector, strongly overestimates the experimental data point. These observations indicate  that new possibilities 
need to be searched for in future research with the aim of achieving a more realistic description of the level splittings. In first place we expect relevance  of  kinematic level splittings as they appear in a Klein-Gordon equation 
with the potential under discussion.
Finally, for a vanishing $d$ value, a finite  number of states belonging to one and the same  ``principal quantum number,''  $N=\ell +n +1$, were predicted, and perfect degeneracies among couples of the type, $2S-1P$, $3S-2P-1D$ etc 
with  $N=2$, and $N=3$, respectively, showed up. 
Without entering into technical details, such occurs because the Schr\"odinger Hamiltonian with  our potential for $d=0$, i.e. ${\mathcal H}(r/a)$ in (\ref{Rania}) for $d=0$, is conformally symmetric due to the circumstance that it is intertwined with the Casimir invariant, ${\mathcal K}^2$, of the rotational group in four dimensions, $SO(4)$, the maximal compact subgroup of the conformal group $SO(2,4)$. Indeed, in this case, the parameter $a$ acquires meaning of radius of a three dimensional hyper-sphere, $S^3$, whose isometry is that very same $SO(4)$. Then, the $(r/a)$ variable acquires meaning of the second polar angle, here denoted by $\chi$, and used in the  parametrization of  $S^3$. The first polar angle, $\theta$, and the azimuthal angle, $\varphi$ are same as in the standard spherical harmonics.  The form of this intertwinement depends on the aforementioned angles and is most simple for the case of the ground state where one can directly calculate that the following relation holds valid,
\begin{eqnarray}
e^{\frac{V_0}{2}\chi}{\mathcal H}(\chi) |_{d=0}&=&
\left[ {\mathcal K}^2(\chi) +\frac{V_0^2}{4}\right] e^{\frac{V_0}{2}\chi} , \quad  \chi:=\frac{r}{a},\nonumber\\
{\mathcal K}^2(\chi)&=&-
\frac{1}{\sin^2\chi}\frac{\partial}{\partial \chi}
\sin^2\chi \frac{\partial }{\partial \chi} +\frac{\ell (\ell +1)}{\sin^2\chi}.
\label{intertwinging}
\end{eqnarray}
In other words, upon changing the wave function according to,
$u_{n\ell}(r/a)/\sin (r/a)=\psi (r/a)$, amounts to  quantum motion on the three-dimensional hyper-sphere, $S^3$. There, the $\ell (\ell +1)\csc^2 (r/a)$ term acquires meaning of the centrifugal barrier on this manifold, while the cotangent function provides a harmonic solution to the $S^3$ Laplacian, pretty much as the $1/r$ function provides a harmonic solution to the regular flat space Laplacian. To the amount the isometric group of $S^3$ is $SO(4)$, the maximal compact group of the conformal group $SO(2,4)$, the quantum motion on $S^3$ perturbed by the harmonic cotangent  potentials alone, necessarily obeys same symmetry as the Laplacian, namely $SO(4)\subset SO(2,4)$ in our case \cite{Kirchbach2016}.  All these considerations follow from the principles underlying the mathematical discipline of potential theory on surfaces. 
 Being con formally symmetric for $d=0$, our potential for $d\not=0$, therefore describes a system whose conformal symmetry has been perturbatively violated through removing the degeneracies in the levels, the $SO(4)$ multiple ts, characterized by the  principle quantum number $N=(n+\ell +1)$. In order to clarify the status of the symmetry, we need to compare the wave functions of the perturbed with the unperturbed case and  figure out whether or not the degeneracy removal is accompanied by mixing of states from different multiplets (levels). When such mixing does not happen, then the symmetry is preserved as 'dynamical'' \cite{Gilmore}.

\begin{figure}
\centering
\subfigure{\includegraphics[scale=0.80]{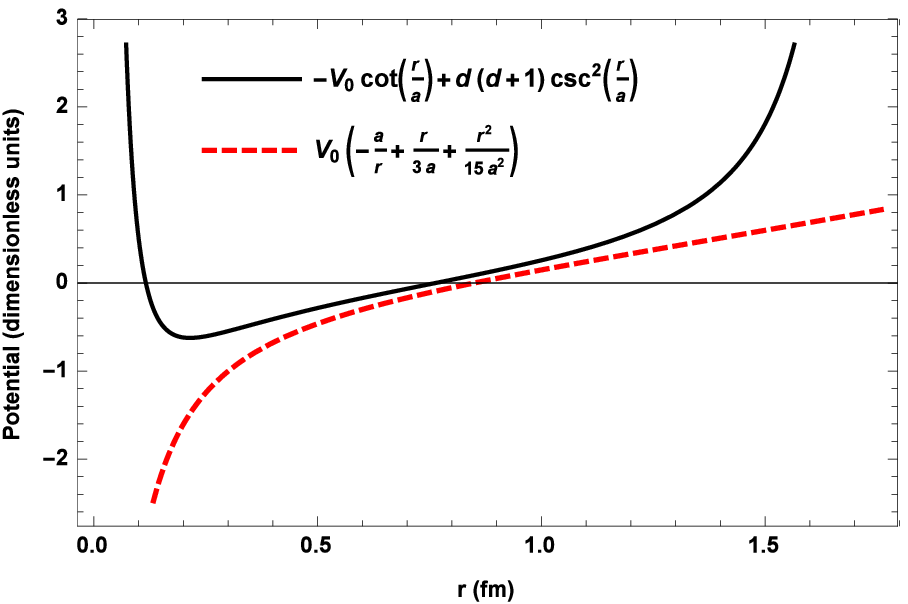}}
\subfigure{\includegraphics[scale=0.80]{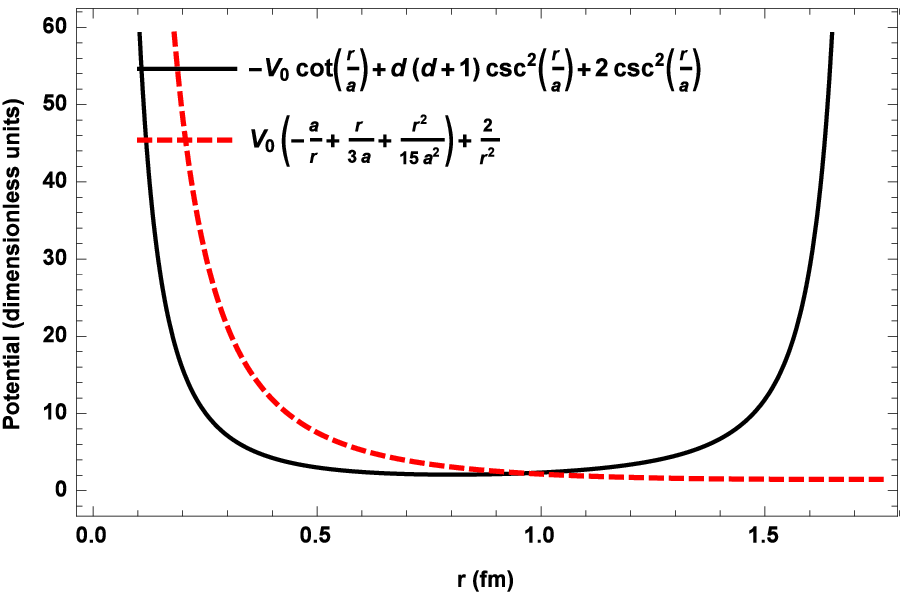}}
\caption{The hard-walled trigonometric Rosen-Morse  potential (solid line) in comparison to the soft-walled Cornell+harmonic oscillator + centrifugal terms (dashed line) 
for the (charmonium) with parameter values of $a$, $d$ as given in Table~(\ref{Table1}) caption, here for $l=0$ (left panel), and  $l=1$ (right panel).}
\label{fig2}
\end{figure}


Using the equation(\ref{wf}), we calculate (up to the normalization constant) the wave functions for $1S$ and $1P$ states for  $c\bar{c}$ and  $b\bar{b}$. 
\begin{eqnarray}
u_{1S}^{c\bar{c}}\left(\frac{r}{a}\right)&=&\e^{-\frac{0.27 r}{a}} \sin^{1.109}\left(\frac{r}{a}\right),\label{wafu1Scharm}\\
u_{1P}^{c\bar{c}}\left(\frac{r}{a}\right)&=&\e^{-\frac{0.147 r}{a}} \sin^{2.039}\left(\frac{r}{a}\right),\label{wafu1Pcharm}\\
u_{2S}^{c\bar{c}}\left(\frac{r}{a}\right)&=&\e^{-\frac{0.142 r}{a}} \sin^{2.109} \left(\frac{r}{a}\right)\left[0.284 
-2.218 \cot\left(\frac{r}{a}\right)\right],\label{wafu2Scharm}\\
u_{1S}^{b\bar{b}}\left(\frac{r}{a}\right)&=&\e^{-\frac{0.265 r}{a}} \sin^{1.131}\left(\frac{r}{a}\right),\label{wafu1Sbottom}\\
u_{1P}^{b\bar{b}}\left(\frac{r}{a}\right)&=&\e^{-\frac{0.146 r}{a}} \sin^{2.048}\left(\frac{r}{a}\right),\label{wafu1Pbottom}\\
u_{2S}^{b\bar{b}}\left(\frac{r}{a}\right)&=&\e^{-\frac{0.141 r}{a}}\sin^{2.131} \left(\frac{r}{a}\right)\left[0.282
-2.26 \cot\left(\frac{r}{a}\right)\right] .\label{wafu2Sbottom}
\end{eqnarray}
The expressions in the brackets are  Romanovski polynomials of first order.
The normalization constant $N_{n\ell}$ for a certain state can be calculated using the condition
\begin{equation}
N_{n\ell}^2\int_{0}^{\pi}|u_{n\ell}\left(\frac{r}{a}\right)|^2 4\pi d\left(\frac{r}{a}\right) =1,
\label{normalize}
\end{equation} 
where $4\pi$ is the integration over the solid angle. With the aim to prove the symmetry status according to the criteria discussed above, we  pick up as an illustrative example the $1S$ wave function 
and factorize its conformally symmetric part corresponding to $d=0$ finding,
\begin{eqnarray}
u_{1S}^{c\bar{c}}\left(\frac{r}{a}\right)&=&\left(\e^{-\frac{0.30 r}{a}} \sin^{0.109}\left(\frac{r}{a}\right)\right)\e^{0.03\frac{r}{a}} \sin\left(\frac{r}{a}\right),\label{wafu1S-2}\\
u_{1S}^{b\bar{b}}\left(\frac{r}{a}\right)&=&\left(\e^{-\frac{0.30 r}{a}} \sin^{0.131}\left(\frac{r}{a}\right)\right)\e^{0.034\frac{r}{a}} \sin\left(\frac{r}{a}\right).
\label{wafu2S-2}
\end{eqnarray}
We observe that the $d$-term in (\ref{RM1}) solely modulates the shape of the conformally symmetric wave function, given by the expression in the round brackets, through the factor remaining outside, but does not provoke any  mixing with the $2S$ state in (\ref{wafu2Scharm}). Similar analyses can be performed with the remaining wave functions. In this manner,the conformal symmetry still maintains its viability in the heavy flavor sectors as a dynamical symmetry \cite{Gilmore}.
In Figures~(\ref{fig3}), we plotted the normalized reduced probability density $|u_{n\ell}(r/a)|^2$ for various states. It can be seen that the curves end in zero at the boundaries $r\in[0,a\pi]$. The smallness of the  $a$ parameter is indicative of a strong spatial localization of the $q-\bar q$ system. As a comparison, we also plotted the spherical Bessel functions,
the ground state solutions to an other hard-walled potential, the infinite spherical ``square'' potential. The node number  $n$  is same in both cases, as it should be. \\

\begin{figure}[!htbp]
\centering
 \subfigure{\includegraphics[scale=0.50]{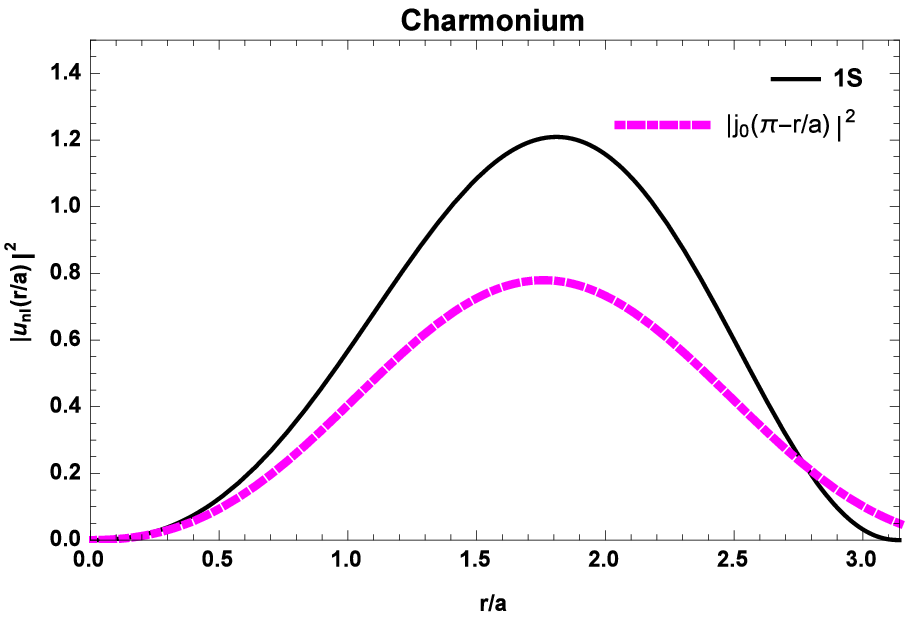}}
 \subfigure{\includegraphics[scale=0.50]{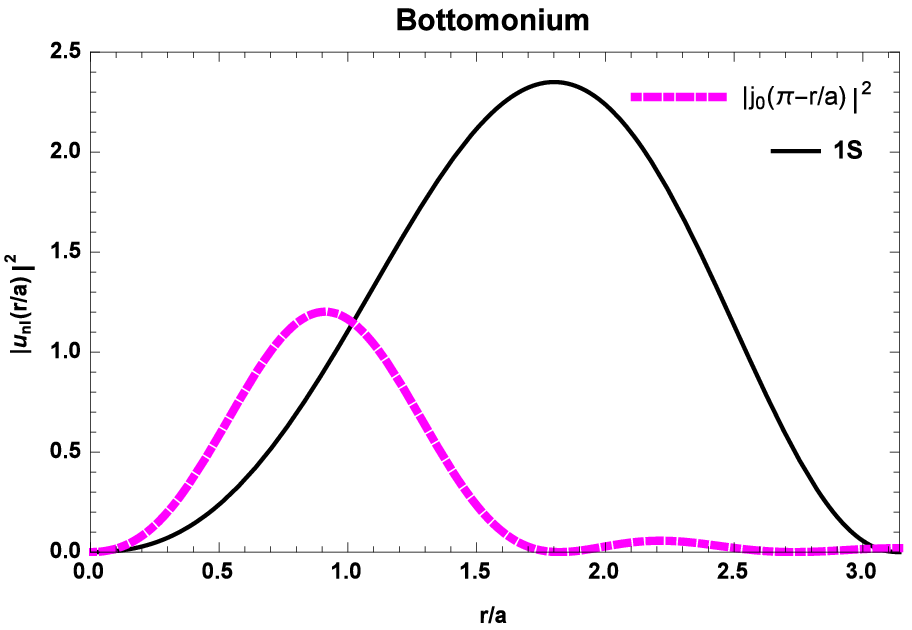}}
\subfigure{\includegraphics[scale=0.50]{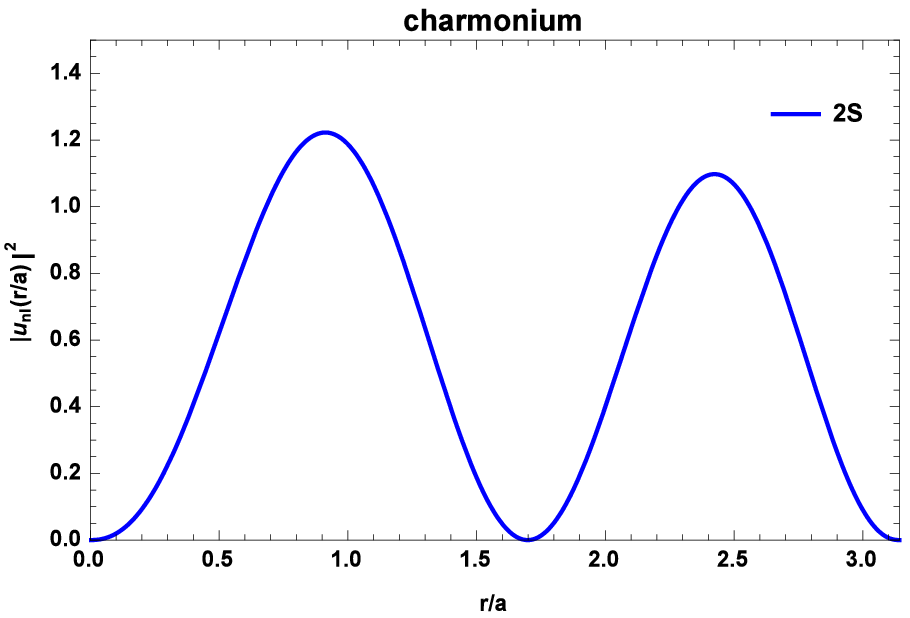}}
 \subfigure{\includegraphics[scale=0.50]{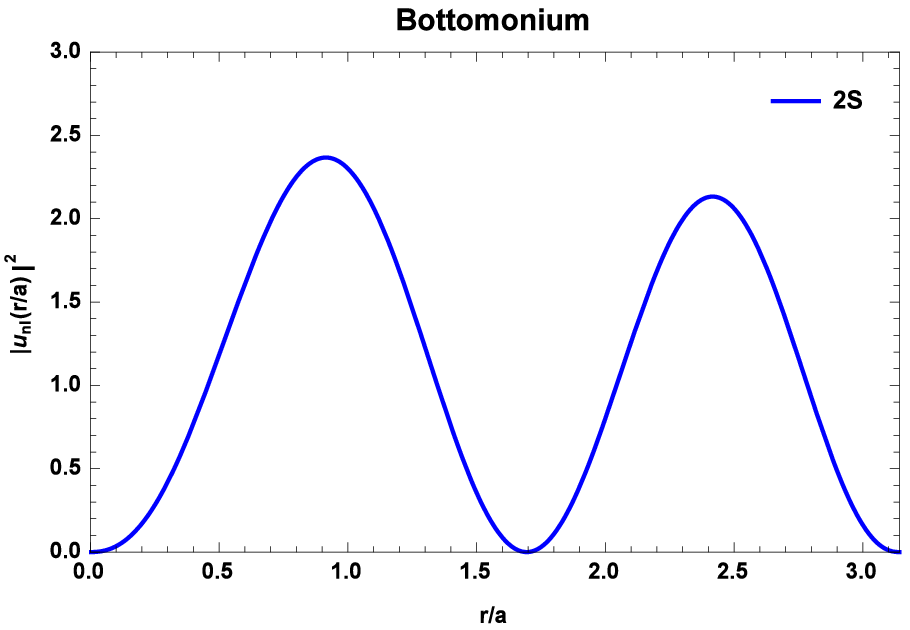}}
 \subfigure{\includegraphics[scale=0.50]{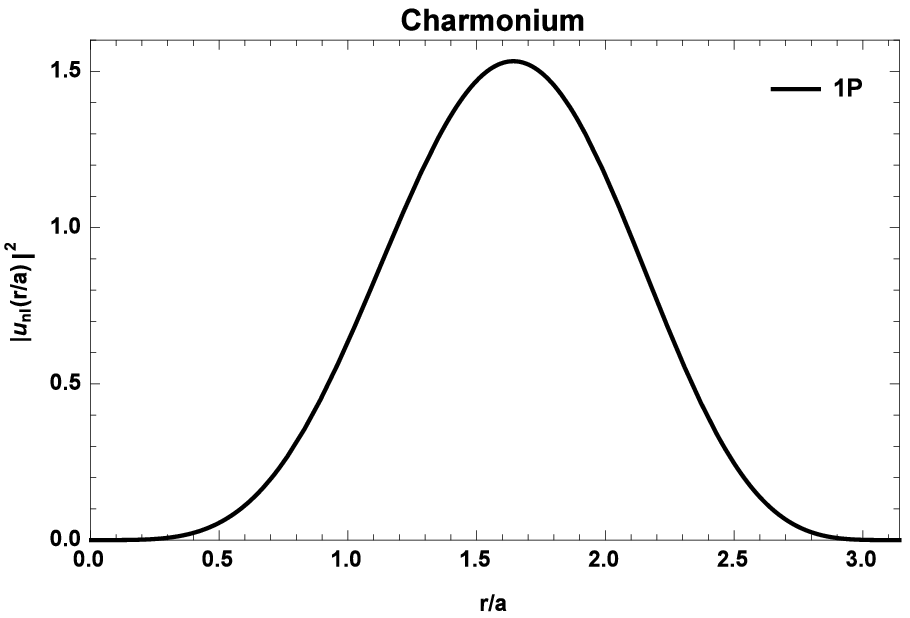}}
 \subfigure{\includegraphics[scale=0.50]{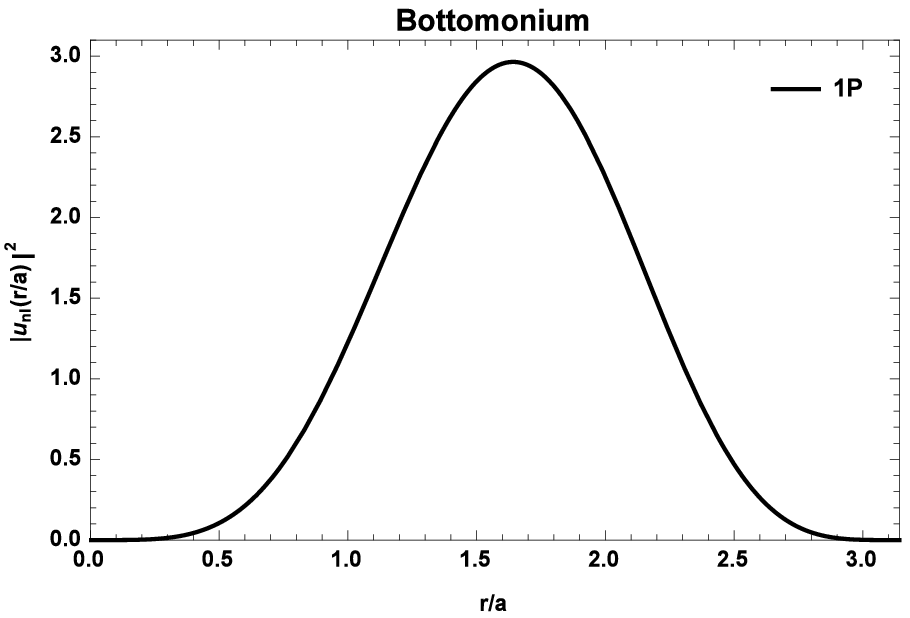}}
\subfigure{\includegraphics[scale=0.50]{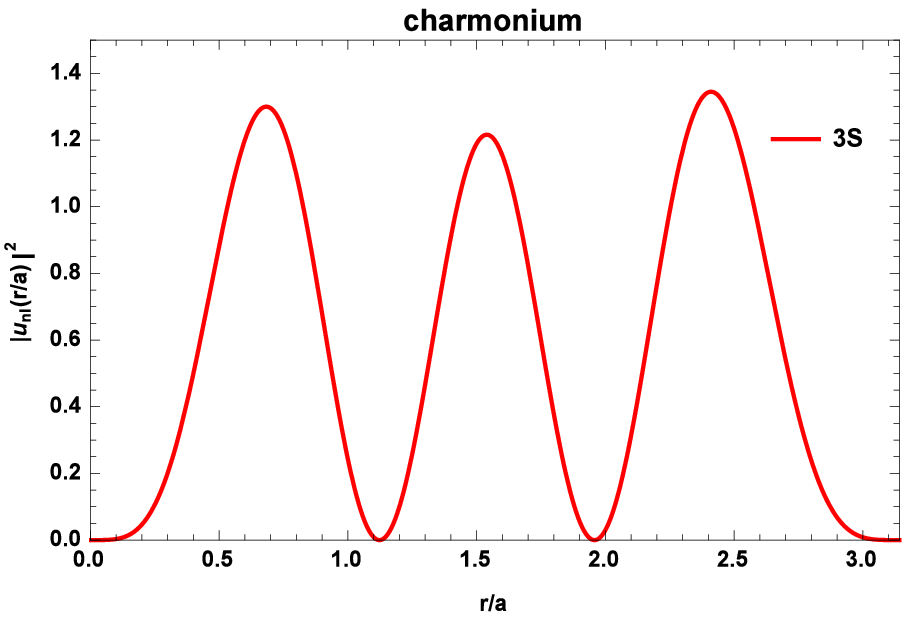}}
 \subfigure{\includegraphics[scale=0.50]{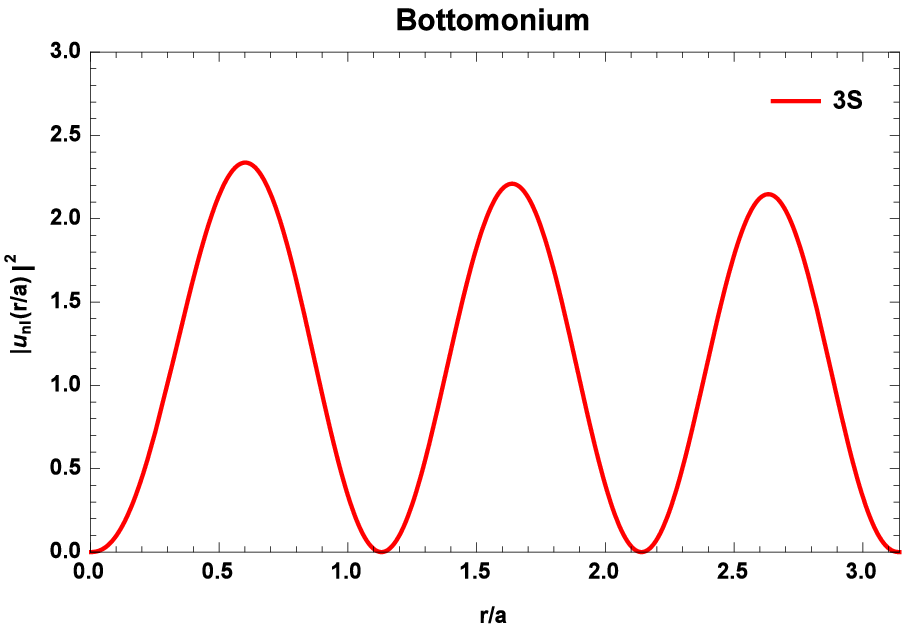}}
\subfigure{\includegraphics[scale=0.50]{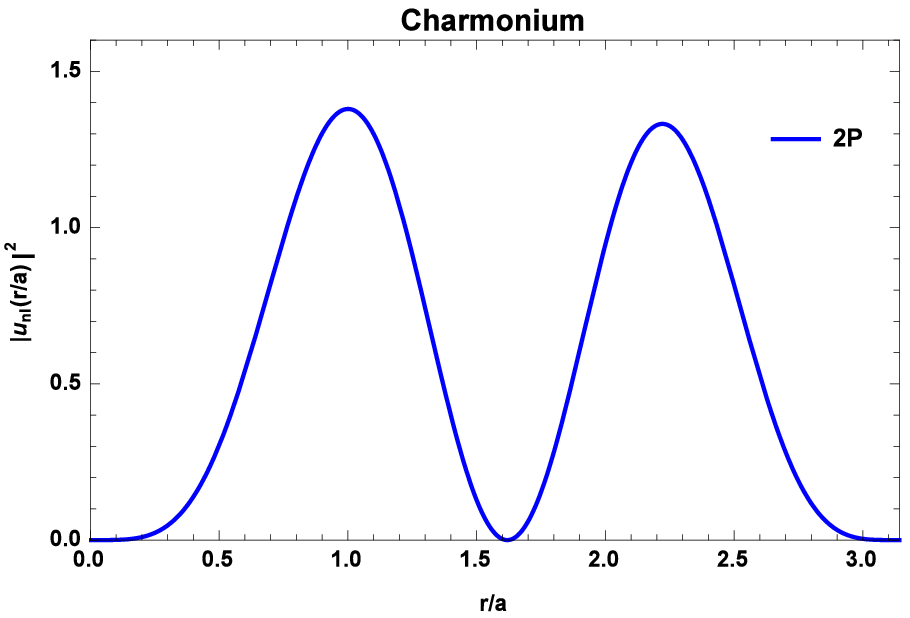}}
 \subfigure{\includegraphics[scale=0.50]{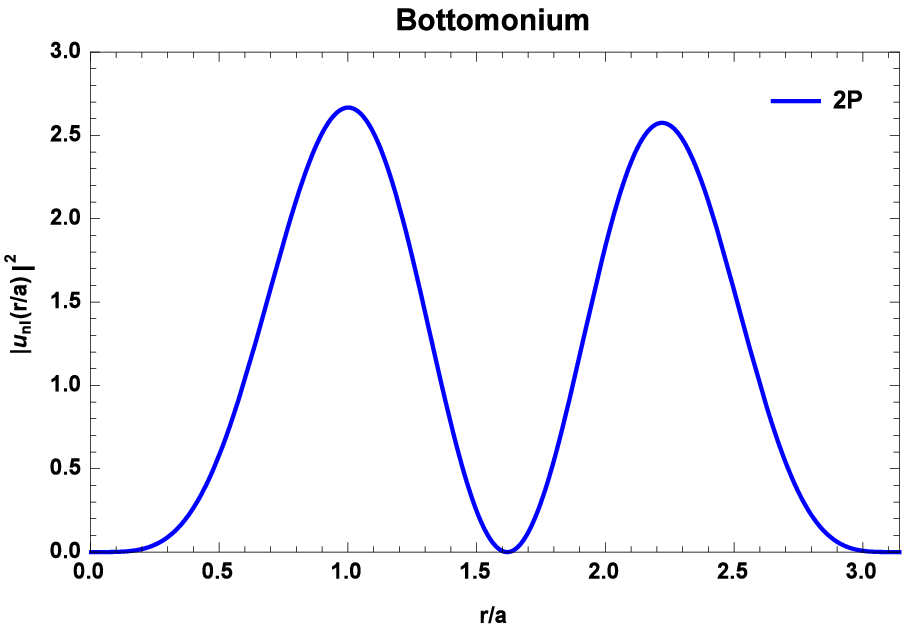}}
\subfigure{\includegraphics[scale=0.50]{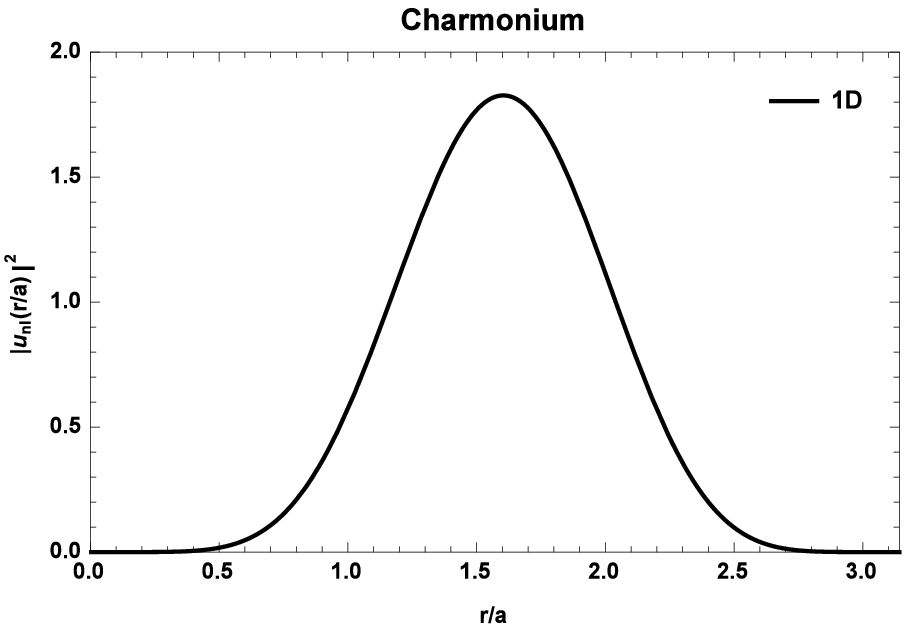}}
 \subfigure{\includegraphics[scale=0.50]{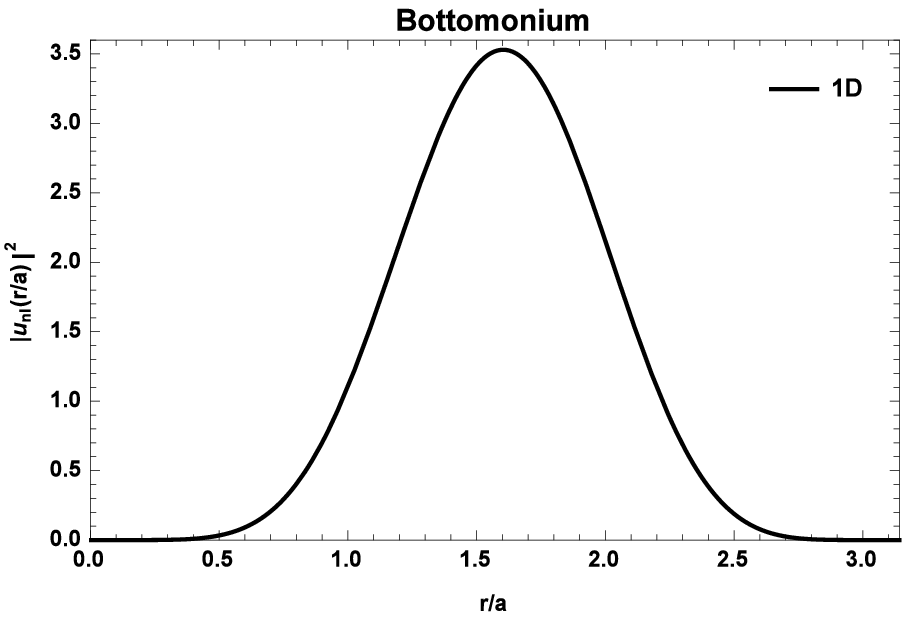}}
 	\caption{The normalized charmonium (left) and bottomonium (right) radial probabilities $\vert u_{n\ell}(r/a)\vert^2$ corresponding from top to bottom to $1S$, $2S$, $1P$,$3S$, $2P$ and $1D$ states, respectively. As a comparison,
the spherical Bessel function $\vert j_0(\pi-r/a)\vert^2$, the ground state solution of the infinite spherical potential,  has been plotted along with the 1S states.}
	\label{fig3}
\end{figure}

\subsection{Root mean square radii}
 Another quantity of physical interest is  the root mean square radius (r.m.s.) $\left\langle r^2 \right\rangle^{1/2}$ for a quark and anti-quark bound state. It it defined as
\begin{equation}
\label{333}
\left\langle r^2 \right\rangle^{1/2}=a\left[\int_{0}^{\pi}|u_{nl}\left(\frac{r}{a}\right)|^2 \left(\frac{r}{a}\right)^2 d\left(\frac{r}{a}\right)\right]^{1/2}.
\end{equation} 
The calculated  $\left\langle r^2 \right\rangle^{1/2}$ values for several quarkonium states, using the potential parameters from the mass fit, are summarized in Table~\ref{Table3}. The results show little  variation of this quantity over different states. We also noticed from Table~\ref{Table3} that the charm-to bottom root mean square radii 
are in a ratio of, 
\begin{equation}
\left\langle r^2 \right\rangle^{1/2}_{c\bar{c}}:\left\langle r^2 \right\rangle^{1/2}_{b\bar{b}}\approx 2:1.
\label{rms_ratio}
\end{equation}
 
In order to get a better insight into  the effect of $d$ on the r.m.s., we 
consider as a particular case the  $1S$  states and use  in (\ref{333}) the small angle approximation for the sine function given by, 
$\sin^\nu (r/a)\approx(r/a)^\nu$. In addition, we extend the integration to the mathematically allowed infinity, with the aim of obtaining an expression in closed form. In so doing, we arrive at the following  result,
\begin{eqnarray}
\label{311}
	\left\langle r^2 \right\rangle^{1/2} &\approx& \int_{0}^{\infty}dr  N_{00}^2 r^2 \left(\frac{r}{a}\right)^{2-2 d} \exp \left(-\frac{2 r V_0}{2 a (d+1)}\right) \nonumber\\
	&=&N_{00}^2\left(\frac{1}{a}\right)^{2-2 d} \Gamma (5-2 d) \left(\frac{V_0}{a d+a}\right)^{2 d-5}.
	\end{eqnarray}
Along same lines, the normalization constant $N_{00}$ defined by Eq.~(\ref{normalize}), now calculates as,
\begin{equation}
\label{33397}
N_{00} \approx \left[\left(\frac{1}{a}\right)^{2-2 d} \Gamma (3-2 d) \left(\frac{V_0}{a d+a}\right)^{2 d-3}\right]^{-1/2}
\end{equation} 
Substituting this in Eq.~(\ref{311}), and then expanding around $d=0$ amounts to,
\begin{equation}
\label{323}
\left\langle r^2 \right\rangle=\frac{3 a^2}{\pi  V_0^2}+\frac{5 a^2 d}{2 \pi  V_0^2}-\frac{3 a^2 d^2}{\pi  V_0^2}.
\end{equation} 
 The latter expression, and for the  parameters in  Tables~\ref{Table1}-\ref{Table2}, leads to $\left\langle r^2 \right\rangle^{1/2}=0.947$ fm for $c\bar{c}$ and $0.494$ fm for $b\bar{b}$, respectively. Their ratio amounts to,
\begin{equation}
\label{3339}
\frac{\left\langle r^2 \right\rangle^{1/2}_{c\bar{c}}}{\left\langle r^2 \right\rangle^{1/2}_{b\bar{b}}}= 1.92,
\end{equation} 
a value that is pretty close to the experimental one given in (\ref{rms_ratio}). Because of the smallness of the $d$ parameter, the equation (\ref{3339}) in combination with (\ref{323}) clearly shows that the major part of the ratio in (\ref{rms_ratio}) is predominantly  provided by the ratio of the corresponding $a$ parameters, which is $a|_{c\bar c}:a|_{b\bar b}=1.93$. We interpret this result as  a stronger localization of the $b\bar b$ system relative to the $c\bar c$ system.
 
It is important to emphasize  that mass spectra and r.m.s. radii have been calculated by one and the same parameter set $a,d,V_0, \mu$.  
\begin{table}[!htbp] \caption{The $\left\langle r^2 \right\rangle^{1/2}$ (in fm) of $c\bar{c}$ and $b\bar{b}$ for various states. Other theoretical predictions are taken from \cite{Gonzalez09}}. 
\centering 
\begin{tabular}{c|cccc}
\hline
$nL$ & This work $c\bar{c}$ & From \cite{Gonzalez09} $c\bar{c}$ & This work $b\bar{b}$ & From \cite{Gonzalez09} $b\bar{b}$ \\ \hline  
1S &0.848&---&0.438&0.2\\
1P &0.890&0.7&0.457&0.4\\
1D &0.889&1.0&0.460&0.4\\
2S &1.001&0.9&0.512&0.5\\
3S &1.034&1.4&0.522&0.8\\
3P &0.959&1.8&0.506&1.0\\ 
2D &0.916&1.5&0.485&0.8\\  
1F &0.888&---&0.461&---\\ 
\hline
\end{tabular}
\label{Table3}
\end{table}
\begin{figure}[!tbp]
\centering
\subfigure{\includegraphics[scale=0.80]{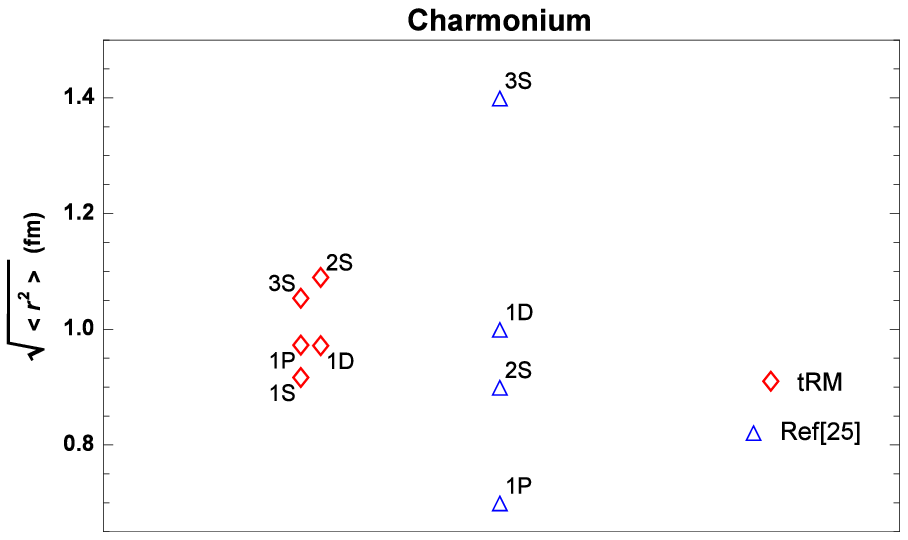}}
\subfigure{\includegraphics[scale=0.80]{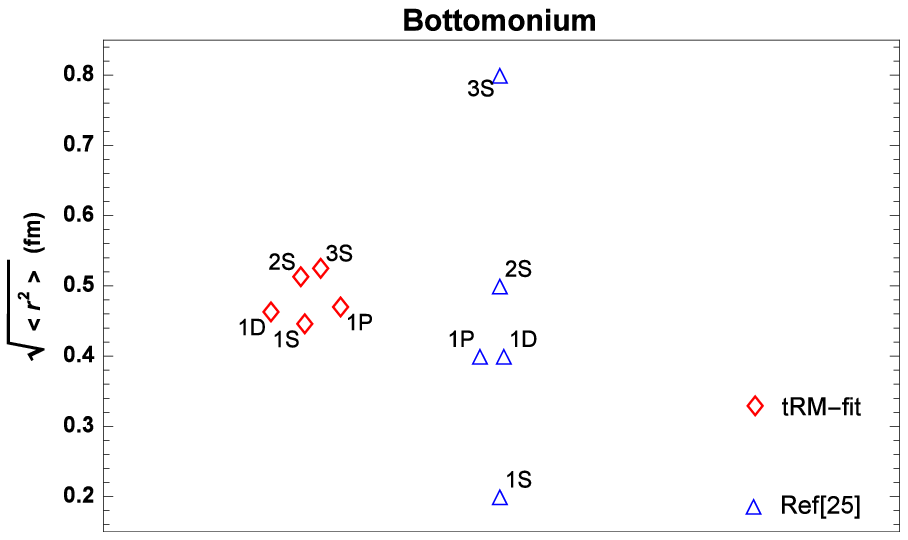}}
\caption{The root mean square radius for charmonium (left panel), and  bottomonium (right panel). Our results are compared with other data taken from \cite{Gonzalez09}.}
\label{fig4}
\end{figure}

\section{Conclusions}
In this work, we obtained the energy eigenvalues and the corresponding eigenfunctions for the hard-wall trigonometric Rosen-Morse potential in (\ref{cot}) using the Nikiforov-Uvarov method.  The expressions in closed forms for wave functions for different states have been explicitly given in the eqs.~(\ref{wafu1Scharm}-\ref{wafu2Sbottom}).  The corresponding reduced probability densities have been  plotted in Fig.~\ref{fig3}, and compared with the spherical Bessel function $j_0$, the ground state solutions that appear in the hard sphere method. As discussed after Eq.~(\ref{RM1}), the potential of interest allows for an  interpretation as a conformal symmetry perturbing strong interaction, a topic of interest in the heavy flavor sector. The scheme has been applied to the ($c\bar{c})$ and $(b\bar{b}$) quarkonia and the results obtained on their  mass spectra and root mean square
radii  have been compared to  data, finding quite satisfactory  agreements. Mass spectra and root mean square radii have been fairly well described by means of a Hamiltonian of moderately perturbed conformal symmetry, as signaled by the smallness of the symmetry violating parameter $d$, a fact that hints on the  validity of a dynamically realized  conformal symmetry in the heavy flavor sector. The dynamical realization of the  symmetry under discussion allowed us  to predict the quantum numbers appearing in  the spectra and 
the splittings between the states in a level. The method predicted  all correct signs of the splittings among the states in the levels, although strongly underestimated their magnitudes. We expect to find this situation improved  by accounting for  the relativistic kinematic level splittings as they would arise in employing the potential under discussion in a Klein-Gordon equation, a goal for a future research.  We also were able to realistically describe the root mean
 square radii of the $c\bar c$ and $b\bar b$ systems, finding them in a ratio very close to $2:1$, a fact which we interpreted as a stronger localization of the $b\bar b$ mesons relative to the $c\bar c$ mesons. 

   Our major conclusion is that conformal symmetry seems to remain present as a viable dynamical symmetry  in all regimes of QCD. As a reminder, dynamical symmetry realization allows for degeneracy removal inside any one of the multiplets but forbids  mixing among the multiplets \cite{Gilmore}. 
Moreover, our conformal symmetry perturbing  parameter $d$ in (\ref{RM1}) features a realistic mass dependence in so far as it  increased in the heavier bottomonium sector by a factor of about 1.6 relative to  value characterizing the lighter charmonium sector, as it should be. 


\end{document}